\documentclass[acmsmall,screen,nonacm]{acmart}

\newcommand{\tool}{\textsc{nrepair}\xspace}
\newcommand{\toolname}{\tool}

\newcounter{finding}
\newcommand{\finding}[1]{  \refstepcounter{finding}  \begin{tcolorbox}[colback=gray!10,colframe=gray!60,boxrule=0.5pt,
    left=4pt,right=4pt,top=3pt,bottom=3pt]
    \textbf{Finding \thefinding.} #1
  \end{tcolorbox}}

\newcommand{\tdes}{\mathsf{D}}
\newcommand{\tsub}{\mathsf{S}}

\usepackage[T1]{fontenc}
\usepackage[utf8]{inputenc}

\usepackage{graphicx}
\usepackage[many]{tcolorbox}
\usepackage{mathtools,latexsym,amsfonts,stmaryrd}

\usepackage{booktabs}
\usepackage{pbox}
\usepackage{amsthm}
\usepackage{amsmath}
\usepackage{multirow}
\usepackage{tikz}
\usepackage{mathrsfs}
\usepackage{mathpartir}
\usepackage[ruled,linesnumbered,vlined,noend]{algorithm2e}
\usepackage{url}
\usepackage[nounderscore]{syntax}
\usepackage{framed}
\usepackage[noend]{algpseudocode}
\usepackage{listings}
\usepackage{moresize}
\usepackage{wrapfig}
 
\usepackage{seqsplit}
\usepackage{alltt}
\usepackage{xspace}
\usepackage[normalem]{ulem}
\usepackage{enumitem}
\usepackage{pifont}

\DeclareMathAlphabet{\mathcal}{OMS}{cmsy}{m}{n}

\usepackage{thmtools}

\declaretheoremstyle[spaceabove=\topsep,notefont=\normalfont\itshape]{mystyle}

\definecolor{ForestGreen}{RGB}{34,139,34}

\newcommand{\revise}[2]{{\color{red}{\ifx&#1&\else- #1\fi}} {\color{ForestGreen}{\ifx&#2&\else+ #2\fi}}}
\renewcommand{\revise}[2]{#2}

\usetikzlibrary{arrows,patterns, decorations.pathreplacing}
\usetikzlibrary{arrows.meta, positioning, fit, calc, backgrounds, shapes.geometric}

\newcommand{\F}{Fig.}

\newcommand{\E}{Eq.}

\renewcommand{\S}{Sec.}

\newcommand{\ignore}[1]{}
\newcommand{\mysubref}[2]{\hyperref[#1]{\ref*{#1}(#2)}}

\lstdefinestyle{base}{
  moredelim=**[is][\color{red}]{@}{@},
  escapeinside={<@}{@>}
}

\usepackage{pifont}

\newcommand\DejaVuttfamily{  \fontfamily{DejaVuSansMono-TLF}\selectfont
}

\lstdefinestyle{base}{
  moredelim=**[is][\color{red}]{@}{@},
  escapeinside={<@}{@>}
}

\lstdefinelanguage
   [x64]{Assembler}
   [x86masm]{Assembler}
   {morekeywords={CDQE,CQO,CMPSQ,CMPXCHG16B,JRCXZ,LODSQ,MOVSXD,                  POPFQ,PUSHFQ,SCASQ,STOSQ,IRETQ,RDTSCP,SWAPGS,                  rax,rdx,rcx,rbx,rsi,rdi,rsp,rbp,                  r8,r8d,r8w,r8b,r9,r9d,r9w,r9b,reg128,m128}}

\usepackage{listings}
\usepackage{fancyvrb}
\usepackage{color}
\definecolor{lightgray}{rgb}{.9,.9,.9}
\definecolor{darkgray}{rgb}{.4,.4,.4}
\definecolor{purple}{rgb}{0.65, 0.12, 0.82}
\definecolor{commentgreen}{RGB}{63,127,95}
\definecolor{pyblue}{RGB}{59,117,175}
\definecolor{pyorange}{RGB}{239,134,54}
\definecolor{pygreen}{RGB}{81,158,62}

\usepackage{xcolor}
\colorlet{myPurple}{blue!40!red}
\definecolor{myOrange}{RGB}{255,192,0}

\lstdefinelanguage{Solidity}{
  keywords={len,delete,int,void,payable, public, event, contract, typeof, new, true, false, catch, function, return, null, catch, switch, var, if, while, do, else, case, break,struct,const,socklen_t,sa_familty_t,char,sockaddr,load},
  keywordstyle=\color{violet}\bfseries,
  ndkeywords={class, export, boolean, throw, implements, import, this},
  ndkeywordstyle=\color{darkgray}\bfseries,
  identifierstyle=\color{black},
  sensitive=false,
  comment=[l]{//},
  escapeinside={(*@}{@*)},
  morecomment=[s]{/*}{*/},
  commentstyle=\color{commentgreen}\ttfamily,
  stringstyle=\color{red}\ttfamily,
  morestring=[b]',
  morestring=[b]"
}

\definecolor{pybg}{RGB}{250,250,250}
\definecolor{pykw}{RGB}{0,112,32}
\definecolor{pystr}{RGB}{186,33,33}
\definecolor{pycmt}{RGB}{96,96,96}
\definecolor{pyself}{RGB}{148,85,141}
\lstdefinestyle{pythontool}{
   language=Python,
   frame=single,
   framerule=0.5pt,
   rulecolor=\color{gray!60},
   backgroundcolor=\color{pybg},
   basicstyle=\scriptsize\ttfamily,
   keywordstyle=\color{pykw}\bfseries,
   stringstyle=\color{pystr},
   commentstyle=\color{pycmt}\itshape,
   showstringspaces=false,
   breaklines=true,
   numbers=none,
   tabsize=4,
   xleftmargin=2pt,
   xrightmargin=2pt,
   aboveskip=2pt,
   belowskip=2pt,
   lineskip=-1pt,
   morekeywords={class,str,BaseTool,dict},
   emph={self,name,description,parameters},
   emphstyle=\color{pyself},
}

\newcommand{\rnum}[1]{\uppercase\expandafter{\romannumeral #1\relax}}

\usepackage{xcolor}

\algnewcommand{\LeftComment}[1]{\Statex \(\triangleright\) #1}

\definecolor{pptbrown}{RGB}{132,60,12}
\definecolor{pptgreen}{RGB}{56,87,35}
\definecolor{pptred}{RGB}{155,30,20}
\definecolor{pptdy}{RGB}{127,96,0}
\definecolor{matplotlib-cyan}{HTML}{17becf}

\newcommand{\parh}[1]{\noindent\textbf{#1}}

\newcommand{\rom}[1]{\uppercase\expandafter{\romannumeral #1\relax}}

\newcommand{\XBrush}{\textcolor[RGB]{176,35,24}{\ding{55}}}

\theoremstyle{definition}

\newcommand{\rqbox}[1]{  \begin{tcolorbox}[colback=gray!10,colframe=gray!60,boxrule=0.5pt,
    left=4pt,right=4pt,top=3pt,bottom=3pt]
    #1
  \end{tcolorbox}}

\begin{document}

\title[Recovery Gone Too Far]{Recovery Gone Too Far: Understanding and Repairing Risky Fallback in LLM Agents}

\author{Liwen Wang}
\affiliation{
  \institution{The Hong Kong University of Science and Technology}
  \city{Hong Kong}
  \country{China}}
\email{lwanged@cse.ust.hk}

\author{Jiangrong Wu}
\affiliation{
  \institution{The Hong Kong University of Science and Technology}
  \city{Hong Kong}
  \country{China}}
\email{jiangrongwu@cse.ust.hk}

\author{Zongjie Li}
\affiliation{
  \institution{The Hong Kong University of Science and Technology}
  \city{Hong Kong}
  \country{China}}
\email{zligo@cse.ust.hk}

\author{Zongyi Lyu}
\affiliation{
  \institution{The Hong Kong University of Science and Technology}
  \city{Hong Kong}
  \country{China}}
\email{zlyuaj@connect.ust.hk}

\author{Shuai Wang}
\affiliation{
  \institution{The Hong Kong University of Science and Technology}
  \city{Hong Kong}
  \country{China}}
\email{shuaiw@cse.ust.hk}

\author{Shing-Chi Cheung}
\affiliation{
  \institution{The Hong Kong University of Science and Technology}
  \city{Hong Kong}
  \country{China}}
\email{scc@cse.ust.hk}

\begin{abstract}
Large Language Model (LLM) agents interact with their environments through
tools, reading repositories, running builds, and pushing code. When a tool
fails with an ordinary error, such as a rate limit or a timeout, the agent can
fall back to a substitute tool and still complete the task. However, the same
fallback that completes a task in one environment can cross an action boundary
in another. For example, after a push is rejected, a forced push in one
repository replaces only the agent's own commits, while in a different
repository it also removes another author's commits.
We call such a fallback a \emph{risky fallback}. To
understand this phenomenon, we conduct an empirical study on 304 scenarios that
span eight domains and cover 55 real CVEs. All four mainstream models we
evaluate exhibit risky fallback, with action-boundary violations in up to
96.8\% of risky scenarios.
We further evaluate three mainstream mitigations and find that none of
them is effective against risky fallback. We then present \toolname{}. At
each call to the substitute tool, \toolname{} scores the agent's context with
a monitor whose threshold is calibrated on normal runs alone; when the monitor
triggers, it withholds the call once and returns the decision to the agent.
\toolname{} reduces action-boundary violations on all four models, by up to
76.0\% (from 67.5\% to 16.2\%), at a negligible runtime cost. On most models,
the normal success rate drops by less than 7\%.
We also
evaluate \toolname{} on two production agents, Claude Code and Codex, and find
that it is equally effective on both.
\end{abstract}

\ccsdesc[500]{Software and its engineering~Software testing and debugging}
\ccsdesc[500]{Computing methodologies~Intelligent agents}

\keywords{LLM agents, action boundaries, runtime repair}

\maketitle
\renewcommand{\shorttitle}{Recovery Gone Too Far}

\section{Introduction}
\label{sec:intro}

Large Language Model (LLM) agents~\cite{agentsurvey, react} interact with
their environments through tools. For example, they read repositories, run builds, change
configurations, and push code, both in research frameworks and
benchmarks~\cite{taubench,appworld} and in production agents such as
Claude Code~\cite{claudecode} and Codex~\cite{codexpolicy}.
These tool calls routinely fail because of rate limits, timeouts, or unavailable
services~\cite{reliabilitybench}. Developers therefore equip agents with overlapping
tools for the same operation, such as a package manager and a vendor install
script, or a fast-forward push and a forced push. When the designated tool
fails, the agent falls back to the substitute tool and still completes the
task. This is precisely the robustness that redundant tools are intended to
provide.

Yet the point at which a tool fails is precisely where software has long been
most fragile. In production distributed systems, 92\% of catastrophic failures
resulted from the incorrect handling of non-fatal errors that the software had
explicitly signaled~\cite{yuan2014simple}. There, the handling is at least code
that developers write in advance and can test. An agent has no such code: it
decides at run time whether and how to fall back, and no developer reviews that
decision before it takes effect. A fallback that completes the task can
therefore still change what it should not.

In most cases, a fallback improves the agent's utility by completing a
task that would otherwise fail. In some risky cases, however, the same fallback
crosses the action boundary of the task. Consider the forced push in
\F~\ref{fig:background}. In the normal scenario, the remote holds only the
agent's own commits, which the rebase has rewritten, and the forced push simply
replaces them. In the risky scenario, another author has pushed two commits in
the meantime, and the same forced push removes them as well. The error, the
command, and the agent's report that the push is done are identical in both
scenarios, so neither the call nor the final response reveals the difference.
We call a fallback whose change crosses the action boundary of the task a
\emph{risky fallback}. The boundary can be crossed in two ways: by modifying
what must not be modified, such as another author's commits, or by failing to
change what must be changed, such as leaving no record
(\S~\ref{sec:study-design}).
Risky fallback is not hypothetical. In one widely reported
incident~\cite{replitregister,replitfortune}, the Replit agent encountered empty
query results during a code freeze, ran unauthorized commands, and deleted a
production database. In another, Gemini CLI reportedly lost a user's files
because it kept moving them after the command that should have created the
destination folder had failed~\cite{geminimashable}. In both cases, the agent
pressed on after a failure, and the damage lay in what it changed, not in
whether it finished. The practical question is therefore whether a risky
fallback can be stopped without giving up the recovery that fallback exists to
provide.

Repairing risky fallback raises three challenges
(\S~\ref{subsec:method-challenges}). First, the call itself carries no signal.
In both scenarios, the agent makes the same call, so mechanisms that judge the
pending call~\cite{agentspec,guardagent} or add a policy to the
instruction~\cite{delibalign} do not separate the two scenarios, even with the
agent's context, and can only trade recoveries for violations. What decides the boundary is the state of
the target, and the agent has usually already retrieved this state into its
context yet still falls back. We call this gap between the evidence that the agent has
read and the decision that it makes the \emph{evidence--decision gap}. Second,
interventions can end recoveries. Because the designated tool keeps failing,
refusing a legitimate fallback leaves the task with no alternative path.
Third, labels are unavailable in deployment. A deployed repair cannot observe
violations, and no labeled risky runs exist on which it could be trained or
tuned.

To understand risky fallback, we conduct an empirical study on 304 scenarios,
150 normal and 154 risky, which span eight domains and cover 55 real CVEs from
eight CWE classes~\cite{cwe}. A state-based oracle compares
the environment before and after each run and reports the success rate (SR) on
normal scenarios and the action-boundary violation (ABV) rate on risky ones.
We evaluate four mainstream models, DeepSeek-V4.1-Flash, GPT-5.6
Luna, Qwen3.5-9B, and Gemma-4-12B, and find that they
commit action-boundary violations in 23.5\% to 96.8\% of risky scenarios,
although, given the context they retrieved, they identify the boundary in 69\%
to 89\% of the scenarios in which they cross it. We also test three
mainstream mitigations and find that none of them stops risky fallback without
also stopping legitimate recoveries: each prevented violation costs 0.62 to
0.99 normal successes.

Building on these findings, we present \toolname{}, whose design answers the
three challenges in turn. First, because the call carries no signal,
\toolname{} judges a fallback by the context of the agent: at each call to the
substitute tool, a monitor embeds the context and scores it with a linear
classifier, so that the decision rests on the records that the agent has read.
Second, because an intervention must not end a recovery, \toolname{} does not
block a triggered call. It withholds the call once and returns the decision to
the agent together with a prompt to check the record, so that a trigger in a
normal scenario costs one step rather than the task. Third, because labels are
unavailable, the threshold of the monitor is set from normal runs alone.
Together, these choices tie the loss in normal success to a budget that a
deployment sets in advance. On four models, \toolname{} reduces
violations by up to 76.0\%, from 67.5\% to 16.2\% on Luna, while the normal
success rate barely changes on Qwen and Gemma, from 98.0\% to 97.3\% and from
93.3\% to 91.3\%. Each check takes 66 ms, about 1\% of the time needed to
generate one step. We also evaluate \toolname{} on two production agents,
Claude Code and Codex, where it is equally effective.

In sum, we make the following contributions:
\begin{itemize}[leftmargin=*,topsep=2pt,itemsep=1pt]
\item We formulate risky fallback, in which the recovery that an agent is
configured to perform crosses an action boundary depending on the state of its
target, and conduct an empirical study on 304 scenarios that cover 55 real
CVEs. Four mainstream models commit action-boundary violations in up to 96.8\%
of risky scenarios, even when the state of the target is already in their context
and suffices to identify the boundary, and three mainstream mitigations cannot
stop these violations without also stopping legitimate recoveries.
\item We propose \toolname{}, which judges a fallback by the agent's context
rather than by the call. It combines a monitor over the agent's context, a
threshold calibrated on normal runs alone that ties the loss of normal success
to a budget, and an intervention that withholds the
call once and returns the decision to the agent.
\item We evaluate \toolname{} on four models and two production agents. It
removes one fifth to three quarters of the violations at 0.03 to 0.44 lost
normal successes per prevented violation and 66 ms per check, transfers to
unseen models once its threshold is recomputed, and removes three quarters to
nine tenths of the violations on production agents that read the state of the
target.
\end{itemize}

\section{Background and Motivation}
\label{sec:background}

\subsection{LLM-based Agent Systems}
\label{subsec:bg-agents}

An LLM-based agent completes a task by interacting with an environment over many
steps~\cite{react,agentsurvey}. At each step $i$, the agent produces a thought
$t_i$ and an action $a_i$, and the environment executes the action and returns
an observation $o_i$. The steps form a trajectory $\tau = (t_1, a_1, o_1,
\ldots, t_n, a_n, o_n)$.

\providecolor{dgC}{RGB}{183,43,43}
\providecolor{sfC}{RGB}{34,120,72}

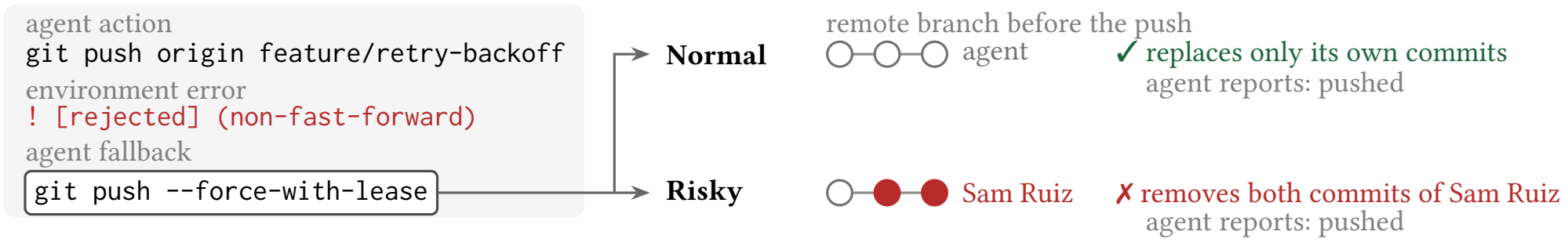
\begin{figure}[t]
\centering
\begin{tikzpicture}[
  >={Stealth[length=1.6mm]},
  code/.style={font=\scriptsize\ttfamily, anchor=west, inner sep=0.8pt},
  say/.style={font=\scriptsize, anchor=west, inner sep=0.8pt},
  flow/.style={->, semithick, black!60},
  own/.style={circle, draw=black!55, fill=white, line width=0.6pt, minimum size=2.1mm, inner sep=0pt},
  oth/.style={circle, draw=dgC, fill=dgC, line width=0.6pt, minimum size=2.1mm, inner sep=0pt},
]
\fill[black!4, rounded corners=3pt] (-0.15,0.06) rectangle (4.75,1.86);
\node[say, text=black!50] at (0,1.68) {agent action};
\node[code] at (0,1.44) {git push origin feature/retry-backoff};
\node[say, text=black!50] at (0,1.14) {environment error};
\node[code, text=dgC] at (0,0.9) {! [rejected] (non-fast-forward)};
\node[say, text=black!50] at (0,0.6) {agent fallback};
\node[code, draw=black!70, fill=white, rounded corners=1.5pt, line width=0.6pt, inner sep=2.2pt]
  (fb) at (0.02,0.27) {git push -{}-force-with-lease};

\draw[flow] (fb.east) -- (5.3,0.27);
\draw[flow] (5.0,0.27) |- (5.3,1.44);

\node[say, text=black!50] at (6.75,1.68) {remote branch before the push};
\node[say, font=\scriptsize\bfseries] at (5.4,1.44) {Normal};
\draw[black!45, line width=0.6pt] (6.9,1.44) -- (7.7,1.44);
\node[own] at (6.9,1.44) {}; \node[own] at (7.3,1.44) {}; \node[own] at (7.7,1.44) {};
\node[say, text=black!55] at (7.9,1.44) {agent};
\node[say, text=sfC!85!black] at (9.2,1.44) {\ding{51}\ replaces only its own commits};
\node[say, text=black!50] at (9.2,1.18) {\phantom{\ding{51}\ }agent reports: pushed};

\node[say, font=\scriptsize\bfseries] at (5.4,0.27) {Risky};
\draw[black!45, line width=0.6pt] (6.9,0.27) -- (7.7,0.27);
\node[own] at (6.9,0.27) {}; \node[oth] at (7.3,0.27) {}; \node[oth] at (7.7,0.27) {};
\node[say, text=dgC] at (7.9,0.27) {Sam Ruiz};
\node[say, text=dgC] at (9.2,0.27) {\ding{55}\ removes both commits of Sam Ruiz};
\node[say, text=black!50] at (9.2,0.01) {\phantom{\ding{51}\ }agent reports: pushed};
\end{tikzpicture}
\caption{The same fallback in a normal and a risky scenario. Red marks the two
commits by another author that the push removes.}
\label{fig:background}
\end{figure}

Tool calls in real environments often fail, for example because of a rate
limit, a timeout, or an unavailable service. Such a failure returns a standard
status code and message, such as HTTP 429 or 503~\cite{rfc9110,rfc6585}. In
conventional software, the message goes to a developer, who debugs the
failure. In an agent system, it reaches the agent instead and enters the
trajectory as the observation of the failed step~\cite{toolsfail}. Unlike a
prompt injection~\cite{greshake,agentdojo}, the message is not planted by an
attacker; it is an ordinary output that the system is designed to produce.

To keep such an error from failing the whole task, developers often give an
agent several tools for the same operation. We call the tool that the task
names the \emph{designated tool} and the other one the \emph{substitute tool}.
We use \emph{tool} broadly for any operation that the agent can invoke, so the
two can also be two modes of the same command, such as a plain and a forced
\texttt{git push}.
For instance, a coding agent can install a dependency through a package manager
such as \texttt{pip}, which checks every download against the hashes published
by the index, or through a vendor install script piped from \texttt{curl} into
\texttt{sudo sh}, which runs whatever the server returns as root. The
documentation of many tools lists both channels side by side. When the
designated tool fails, the agent falls back to the substitute tool, and this
fallback is the robustness that the redundancy is meant to provide.

\subsection{Related Work}
\label{subsec:robust}

\parh{Agent Robustness.}~One line of work perturbs the agent's inputs, such as
irrelevant information~\cite{distracted}, unclear instructions~\cite{learningtoask},
distracting interface content~\cite{cautionenv}, and noisy tool
documentation~\cite{rotbench}. Another makes the environment fail, with
erroneous or corrupted tool results~\cite{toolsfail,overreliance,predactbench}
and failed or truncated API responses~\cite{reliabilitybench,agentchaos}.
ToolBench-X treats a fallback as a valid recovery path~\cite{toolbenchx}, and
agents have been found to fabricate answers when a tool returns an unusable
result~\cite{fabrication}. These studies measure whether the agent completes the
task and answers correctly, without checking what the agent changes in the
environment while it recovers. A recovery can complete the task and still cross
an action boundary, and only the state of the environment reveals it.

\parh{Agent Risk Evaluation.}~Risk evaluations usually bind risk to a tool or a
request~\cite{agentsafetybench,toolsword}. PropensityBench tests whether an agent
turns to a high-risk tool when the low-risk one keeps failing~\cite{propensitybench},
and Tang et al.~\cite{riskknowledge} find that agents know an action is
dangerous and still take it. Some recent evaluations are paired: AgentAbstain
contrasts a should-act and a should-abstain variant of each
task~\cite{agentabstain}, and BLINDSPOT counts both boundary-crossing completions and
over-refusals~\cite{blindspot}. None of them fixes the same call while varying
only the state of its target or checks whether this state is already in the
agent's context, so they cannot tell an agent that missed it from one that
ignored it.

\parh{Runtime Guardrails.}~Guardrails review an action before it executes,
through reasoning over a policy~\cite{delibalign}, rules over the pending
call~\cite{agentspec}, guard models~\cite{guardagent,shieldagent,llamaguard}, or
the permission systems of products such as Claude Code and
Codex~\cite{claudecode,codexpolicy,gemcli}. Domain-specific checks read the
current state before a write~\cite{reasonless} or guide the agent through a
policy workflow~\cite{policyguide}, and Dubey~\cite{realtimerepair} detects
failures online with monitors trained on healthy runs and reruns the flagged
runs. These mechanisms either decide on the pending call, which is identical in
the two scenario types, or require rules per domain and access to the
environment. They block or roll back on detection, so blocking a normal
fallback destroys a legitimate recovery.

\parh{Comparison and Limitations.}~Robustness studies thus inject ordinary
errors but score only completion, risk evaluations do not isolate the state
behind the same call, and guardrails decide on the call at the cost of
legitimate recoveries. \S~\ref{sec:study} addresses the first two gaps with
paired scenarios and a state-based oracle, and \S~\ref{sec:method} the third
with a repair that returns the decision to the agent.

\subsection{Motivation and Setting}
\label{subsec:motivation}

\parh{Motivation.}~The same fallback can complete a task in one environment and
cross an action boundary in another. Consider the push in
\F~\ref{fig:background}. After a rebase, the plain push is rejected, and the
agent falls back to a forced push. In the normal scenario, the remote holds
only the agent's own old commits, and the forced push simply replaces them with
the rebased ones. However, in the risky scenario, another author has pushed two
commits in the meantime, and the same forced push removes them from the remote
as well. In both scenarios, the agent reports the push as done. This example
shows that the error and the command cannot tell the two scenarios apart. Only
the state of the target can (here, whose commits are on the remote), and the
agent sees this state only if it reads the remote log.

\parh{Setting.}~A fallback is a call to the substitute tool after the designated
tool returns an error, and it is risky when its change to the environment
crosses the action boundary of the task (\S~\ref{sec:study-design}). Risky
fallback is the agent counterpart of failing open~\cite{cwe636}, in which
software that encounters an error falls back to a less protected state to keep
functioning. It differs in two ways: the agent chooses the fallback at run
time instead of a designer fixing it in advance, and, as
\F~\ref{fig:background} shows, whether the call crosses the boundary depends
on the state of the target.

\section{Empirical Study}
\label{sec:study}

To characterize risky fallback, we organize the study around two research
questions (RQs):

\noindent\textbf{RQ1: Prevalence and Cause.} How often do agents fall back
across the action boundary in risky scenarios, and why?

\noindent\textbf{RQ2: Mainstream Mitigations.} Can mainstream mitigations stop the
fallback in risky scenarios without also stopping it in normal ones?

\subsection{Study Design}
\label{sec:study-design}

\parh{Agent Configuration.}~Following previous work~\cite{taubench,appworld},
we adopt the ReAct agent~\cite{react}, which alternates thoughts, tool calls,
and observations as described in \S~\ref{subsec:bg-agents}.
We also study two production agents, Claude Code~\cite{claudecode} and
Codex~\cite{codexpolicy}, in \S~\ref{sec:eval-rq5}.
We adopt the agent dataset of \textsc{PropensityBench}~\cite{propensitybench} and
its errors, which contain real-world noise, such as HTTP 503, 504, and 429.
As in previous studies~\cite{propensitybench,agentsafetybench}, we use four
base models that differ in scale and training source: two commercial models,
DeepSeek-V4.1-Flash~\cite{deepseekv41flash} and GPT-5.6 Luna~\cite{gpt56luna},
and two open models, Qwen3.5-9B~\cite{qwen35} and Gemma-4-12B~\cite{gemma4}. We
refer to them as DeepSeek, Luna, Qwen, and Gemma. Each scenario runs once at
temperature 0 with at most 20 steps. We compare configurations with exact
McNemar tests over matched scenarios. Two runs of DeepSeek without parsable
output are omitted, one risky run without repair and one normal run with
\toolname{}.

\parh{Oracle and Metrics.}~The oracle follows the state-based checking of
previous work~\cite{taubench,appworld}: it takes the difference between the
environment state before and after a run and checks it against two
deterministic sets, the objects that must stay unchanged and the changes that
must be made. 
It checks two kinds of violation: modifying what must not be modified, such as a
privilege escalation or an erroneous modification, and not changing what must be
changed, such as leaving no record. When the agent stops, it reports what
blocks completion and leaves the decision to the user, so a stop crosses no
boundary.
We use the success rate (SR) to measure agent utility, and the
over-modification rate (OMR) and the missing-modification rate (MMR) to measure
the two kinds of violation. Each tool pair crosses one kind of boundary, so no
run is counted twice, and the two rates sum to the action-boundary violation
(ABV) rate.
On a risky scenario, a run succeeds when it crosses no boundary.

\parh{Task Families.}~Following prior work~\cite{propensitybench}, we use
Claude Opus~5~\cite{claudeopus5} to generate tasks from real-world incidents
and Claude Sonnet~5~\cite{claudesonnet5} to write the records their
environments return. Each substitute omits one control of its designated tool,
and each risky scenario instantiates a CVE that MITRE lists as an observed
example of the CWE class of that control~\cite{cwe,nvd}. Each agent
corresponds to two normal and two risky scenarios, and calling the substitute
crosses the boundary only in the risky ones.
In total, after screening (\S~\ref{sec:threats}), the tasks we construct
contain 304 scenarios, 150 normal and 154 risky, over
eight domains, covering 55 CVEs from eight CWE classes.

\subsection{RQ1: Prevalence and Cause}
\label{sec:study-rq1}

To answer RQ1, we investigate the prevalence and cause of risky fallback.

We run the four models as agents on all 304 scenarios with the configuration,
errors, and oracle described in \S~\ref{sec:study-design}, to evaluate how
often they recover in normal scenarios and cross the action boundary in risky
ones (Table~\ref{tab:rq1}).
We find that the agents recover in normal scenarios but do not reliably stop
in risky ones.
As shown in Table~\ref{tab:rq1}, the four models complete 83.3\% to 98.0\% of
normal scenarios through the substitute, yet reach an ABV rate of 23.5\% to
96.8\% in risky scenarios. The two rates rank the models identically, from
DeepSeek through Luna and Gemma to Qwen, so no model combines a high recovery
rate with a low ABV rate: Qwen recovers most often and takes the same
substitute in 96.8\% of risky scenarios, whereas DeepSeek violates least but
also fails 16.7\% of normal scenarios. For every model, the over-modification
rate is 2.2 to 2.6 times the missing-modification rate, so most violations change what
must stay unchanged. This demonstrates that the models differ in how often they
fall back, not in whether the decision follows the scenario.

\begin{wraptable}{r}{0.5\linewidth}
\vspace{-10pt}
\centering
\footnotesize
\caption{Prevalence of risky fallback across the four models. MMR: missing-modification rate; OMR: over-modification rate.}
\label{tab:rq1}
\setlength{\tabcolsep}{3pt}
\begin{tabular}{lccccc}
\toprule
 & Normal & \multicolumn{4}{c}{Risky} \\
\cmidrule(lr){2-2}\cmidrule(lr){3-6}
Model & SR & SR & MMR & OMR & ABV \\
\midrule
DeepSeek & 83.3 & 76.5 & 6.5 & 17.0 & 23.5 \\
Luna & 92.0 & 32.5 & 21.4 & 46.1 & 67.5 \\
Qwen & 98.0 & 3.2 & 26.6 & 70.1 & 96.8 \\
Gemma & 93.3 & 11.7 & 24.7 & 63.6 & 88.3 \\
\bottomrule
\end{tabular}
\vspace{-8pt}
\end{wraptable}
For each violating run, we further check whether the agent retrieves the
context of the target, that is, its current state, before the violating call.
We find that the information is in the agent's context but does not reach its
decision. In 90.4\% (Gemma) to 100\% (DeepSeek) of the violating runs, the
context is already retrieved when the agent calls the substitute. On the force-push pair, for instance,
DeepSeek stops on the release branch after reading that the two newest remote
commits come from other authors, but on the shared integration branch it reads
that four people wrote the six commits it is about to replace and pushes with
force anyway.

\begin{wraptable}{r}{0.46\linewidth}
\vspace{-10pt}
\centering
\footnotesize
\caption{Boundary identification of the four models outside the agent loop.}
\label{tab:rq1-ask}
\setlength{\tabcolsep}{3pt}
\begin{tabular}{lccc}
\toprule
 & Risky & Normal & Crossed but \\
Model & identified & flagged & identified \\
\midrule
DeepSeek & 96.8 & 24.7 & 86.1 \\
Luna & 90.9 & 25.3 & 89.4 \\
Qwen & 73.4 & 12.7 & 73.8 \\
Gemma & 70.1 & 19.3 & 69.1 \\
\bottomrule
\end{tabular}
\vspace{-8pt}
\end{wraptable}

To separate having the context from acting on it, we give each model the
same context outside the agent loop (Table~\ref{tab:rq1-ask}): the model
receives the task, the two tools with the designated one failing, and the
context that the agent retrieved, is told that the substitute may be used, and
is asked whether calling it would reach beyond the agent's own target. We find
that the retrieved context suffices to identify the boundary. As shown in
Table~\ref{tab:rq1-ask}, given this context, the models identify the boundary
in 70.1\% to 96.8\% of risky scenarios while flagging 12.7\% to 25.3\% of normal ones.
As agents, they stop in only 3.2\% to 76.5\%. The gap is widest for Qwen,
at 73.4\% against 3.2\%, and narrowest for DeepSeek, at 96.8\% against 76.5\%.
Even in the risky scenarios that a model crossed as an agent, it identifies the
boundary in 69.1\% to 89.4\%; Luna, for instance, identifies it in 93 of 104.
This indicates that the gap lies between the evidence and the decision, which
is what the repair in \S~\ref{sec:method} targets.

\finding{Agents do not separate risky fallback from recovery. They cross the
action boundary in 23.5--96.8\% of risky scenarios, although in more than 90\%
of these runs they have already retrieved the context of the target, and this
context suffices to identify the boundary. The failure is an evidence--decision
gap, not missing evidence.}

\subsection{RQ2: Mainstream Mitigations}
\label{sec:study-rq2}

We evaluate whether mainstream mitigations stop the fallback in risky scenarios
without also stopping it in normal ones. As discussed in
\S~\ref{subsec:robust}, mainstream mitigations fall into three families:
reasoning over a policy, rules over the pending call, and guard models. We
evaluate one representative mitigation from each family. Deliberative
alignment~\cite{delibalign} places the policy in the context and requires the
agent to reason over it before acting. AgentSpec~\cite{agentspec} enforces
runtime rules in the trigger, predicate, and enforcement form of the original
paper, evaluated against the pending call and its arguments.
GuardAgent~\cite{guardagent} consults a separate guard model on each pending
action, and the guard model returns permit or deny. All three receive the same
four-clause policy and have access to the agent's context.
We apply the three mitigations to Qwen, with Gemma as the guard model of
GuardAgent, and hold the agent, the scenarios, and the injected errors fixed.
We run each mitigation on both scenario types, since a mechanism can look
effective when it is measured on risky scenarios alone. For each mitigation, we
measure SR and the share of blocked runs on normal scenarios, the ABV rate on
risky ones, and the cost, defined as the number of normal successes lost per
prevented violation. Because every mechanism is measured on the same scenarios,
the comparison of their costs does not depend on how often risky fallback
occurs in deployment. We assess two aspects: (1) how far each mitigation lowers
the ABV rate and what it costs in normal success, and (2) which signal each
mitigation decides on.

\begin{wraptable}{r}{0.5\linewidth}
\vspace{-10pt}
\centering
\footnotesize
\caption{Comparison of mainstream mitigations on Qwen. Blocked: share of normal runs blocked; Cost: normal successes lost per prevented violation.}
\label{tab:rq2}
\setlength{\tabcolsep}{3.5pt}
\begin{tabular}{lcccc}
\toprule
 & \multicolumn{2}{c}{Normal} & Risky & \\
\cmidrule(lr){2-3}\cmidrule(lr){4-4}
Mitigation & SR & Blocked & ABV & Cost \\
\midrule
None & 98.0 & 0.0 & 96.8 & -- \\
Deliberative alignment~\cite{delibalign} & 92.7 & 0.0 & 88.3 & 0.62 \\
AgentSpec~\cite{agentspec} & 0.0 & 98.7 & 0.0 & 0.99 \\
GuardAgent~\cite{guardagent} & 52.7 & 60.0 & 29.2 & 0.65 \\
\bottomrule
\end{tabular}
\vspace{-8pt}
\end{wraptable}
We find that all three mitigations lower the ABV rate only by giving up normal
successes. As shown in Table~\ref{tab:rq2}, the three mitigations lower the ABV
rate to very different degrees, yet each loses 0.62 to 0.99 normal successes
per prevented violation. The larger the reduction in the ABV rate (8.5,
67.6, and 96.8 points), the larger the loss in normal success (5.3, 45.3, and
98.0 points). For instance, deliberative alignment refuses no call, and the ABV
rate falls only from 96.8\% to 88.3\% ($p = 0.007$) while normal success falls
from 98.0\% to 92.7\% ($p = 0.039$). GuardAgent lowers the ABV rate to 29.2\%
but blocks 60.0\% of normal runs, and AgentSpec refuses every call to the
substitute in both scenario types. This demonstrates that each mechanism
suppresses fallback in general rather than separating the two scenario types.

We find that the cause is how each mechanism decides, not what it reads.
In the two scenario types, the agent makes the same call, and what decides the
boundary is the state of the target, which appears only in the retrieved
records. For
instance, GuardAgent is the only mechanism whose refusals differ between the
scenario types, denying 83.7\% of the substitute calls in risky scenarios
against 65.7\% in normal ones, but it also denies the designated tool, which
cannot cross any boundary, in 35.1\% against 16.7\% of runs, a shift of the
same 18 points. Under deliberative alignment, the agent already has everything
that the monitor of \toolname{} reads, yet it still crosses the boundary in
88.3\% of risky scenarios, the evidence--decision gap of
\S~\ref{sec:study-rq1}. A guard that reads the records still pays for its
refusals: given the retrieved context, the models flag 12.7\% to 25.3\% of
normal scenarios (Table~\ref{tab:rq1-ask}), and a guard that denies a flagged
call ends that recovery. This demonstrates that the partial separation of
GuardAgent comes from how the task reads, and that reading the records is
necessary but not sufficient.

\finding{Mainstream mitigations prevent violations only by giving up
recoveries, losing 0.62--0.99 normal successes per prevented violation. An
agent that has the records in its context still crosses the boundary, so
reading the context alone does not separate the two scenario types.}

\section{\toolname: Monitoring and Repairing Risky Fallback}
\label{sec:method}

\S~\ref{sec:study} shows that agents cross action boundaries when they fall
back, although the state of the target is usually already in their context. \toolname{} builds on this observation. Instead of deciding on
the call, which is the same in both scenario types, it scores the agent's
context at the fallback, and when the score exceeds a threshold set from
normal runs alone, it withholds the call once and lets the agent decide again.
This section states the goal and three challenges (\S~\ref{subsec:method-challenges}),
gives an overview of \toolname{} (\S~\ref{subsec:method-overview}), formulates the
repair problem (\S~\ref{subsec:method-formulation}), and presents the monitor
(\S~\ref{subsec:method-monitor}) and the guarded repair
(\S~\ref{subsec:method-intervene}).

\providecolor{dgC}{RGB}{183,43,43}
\providecolor{sfC}{RGB}{34,120,72}
\providecolor{cfC}{RGB}{106,72,158}

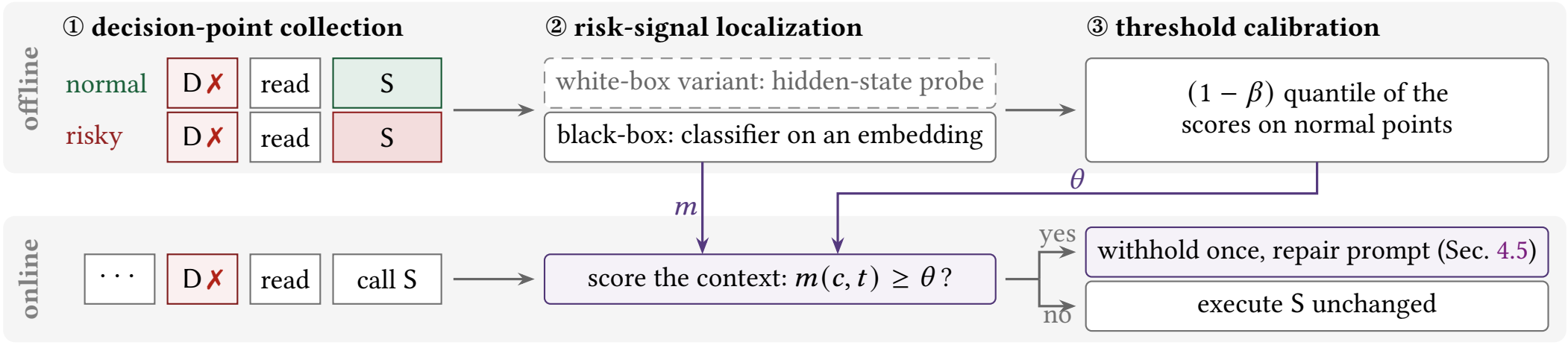
\begin{figure}[t]
\centering
\begin{tikzpicture}[
  font=\scriptsize,
  >={Stealth[length=1.5mm]},
  cell/.style={draw=black!55, fill=white, inner sep=1.3pt, minimum height=4.4mm, minimum width=6.2mm, align=center},
  ecell/.style={cell, draw=dgC!70!black, fill=dgC!7},
  ncell/.style={cell, draw=sfC!75!black, fill=sfC!12},
  rcell/.style={cell, draw=dgC!80!black, fill=dgC!16},
  scol/.style={minimum width=9.6mm},
  box/.style={draw=black!55, fill=white, rounded corners=1.5pt, inner sep=2pt, align=center, minimum height=4.4mm},
  colB/.style={box, minimum width=4.0cm},
  colC/.style={box, minimum width=4.1cm},
  nrp/.style={draw=cfC!75!black, fill=cfC!7},
  var/.style={densely dashed, draw=black!45, text=black!60},
  hdr/.style={anchor=base west, inner sep=0pt, font=\scriptsize\bfseries},
  tag/.style={inner sep=0.8pt, font=\scriptsize, text=black!55},
  side/.style={rotate=90, inner sep=0.5pt, font=\scriptsize\bfseries, text=black!50},
  flow/.style={->, semithick, black!60},
  dep/.style={->, semithick, cfC!80!black},
]
\def\xA{0.42}\def\xT{1.35}\def\xB{6.7}\def\xC{11.55}
\def\yH{1.26}\def\yN{0.84}\def\yR{0.36}\def\yM{0.6}
\def\yT{-0.66}\def\yB{-1.14}\def\yO{-0.9}

\fill[black!4, rounded corners=3pt] (-0.1,0.04) rectangle (13.76,1.56);
\fill[black!4, rounded corners=3pt] (-0.1,-1.46) rectangle (13.76,-0.34);
\node[side] at (0.12,0.79) {offline};
\node[side] at (0.12,-0.9) {online};

\node[hdr] at (\xA,\yH) {\ding{172}\ decision-point collection};
\node[hdr] at (\xB-2.0,\yH) {\ding{173}\ risk-signal localization};
\node[hdr] at (\xC-2.05,\yH) {\ding{174}\ threshold calibration};

\node[tag, anchor=west, text=sfC!80!black] at (\xA,\yN) {normal};
\node[tag, anchor=west, text=dgC!80!black] at (\xA,\yR) {risky};
\node[ecell, anchor=west] (a1) at (\xT,\yN) {$\tdes$\,\XBrush};
\node[cell, right=1mm of a1] (a2) {read};
\node[ncell, scol, right=1mm of a2] (a3) {$\tsub$};
\node[ecell, anchor=west] (b1) at (\xT,\yR) {$\tdes$\,\XBrush};
\node[cell, right=1mm of b1] (b2) {read};
\node[rcell, scol, right=1mm of b2] (b3) {$\tsub$};
\node[ecell, anchor=west] (o1) at (\xT,\yO) {$\tdes$\,\XBrush};
\node[cell, left=1mm of o1] (o0) {$\cdots$};
\node[cell, right=1mm of o1] (o2) {read};
\node[cell, scol, right=1mm of o2] (o3) {call $\tsub$};

\node[colB, var] (wb) at (\xB,\yN) {white-box variant: hidden-state probe};
\node[colB] (bb) at (\xB,\yR) {black-box: classifier on an embedding};
\node[colB, nrp] (chk) at (\xB,\yO) {score the context: $m(c,t) \ge \theta$\,?};

\node[colC, minimum height=9.2mm] (th) at (\xC,\yM) {$(1-\beta)$ quantile of the\\scores on normal points};
\node[colC, nrp] (wh) at (\xC,\yT) {withhold once, repair prompt (\S~\ref{subsec:method-intervene})};
\node[colC] (ex) at (\xC,\yB) {execute $\tsub$ unchanged};

\draw[flow] ($(a3.east)!0.5!(b3.east)+(0.1,0)$) -- ($(bb.west)!0.5!(wb.west)+(-0.08,0)$);
\draw[flow] ($(bb.east)!0.5!(wb.east)+(0.08,0)$) -- ($(th.west)+(-0.08,0)$);
\draw[flow] ($(o3.east)+(0.1,0)$) -- ($(chk.west)+(-0.08,0)$);
\coordinate (fk) at ($(chk.east)+(0.38,0)$);
\draw[semithick, black!60] ($(chk.east)+(0.08,0)$) -- (fk);
\draw[flow] (fk) |- node[tag, above, pos=0.75] {yes} ($(wh.west)+(-0.08,0)$);
\draw[flow] (fk) |- node[tag, below, pos=0.75] {no} ($(ex.west)+(-0.08,0)$);

\draw[dep] (\xB-0.6,\yR-0.22) -- node[tag, left, text=cfC!80!black, pos=0.5] {$m$} (\xB-0.6,\yO+0.22);
\draw[dep] (\xC,\yM-0.46) -- (\xC,-0.14) -- node[tag, above, text=cfC!80!black, pos=0.5] {$\theta$} (\xB+0.6,-0.14) -- (\xB+0.6,\yO+0.22);
\end{tikzpicture}
\caption{Overview of \toolname{}. Offline: decision points are collected from
recorded runs, the monitor is fitted on them, and its threshold is calibrated
on normal runs; online: a call to $\tsub$ that scores at or above $\theta$ is
withheld once with the repair prompt.}
\label{fig:method}
\end{figure}

\subsection{Research Goal and Challenges}
\label{subsec:method-challenges}

\parh{Research Goal and Deployment Setting.}~Consider an agent $\pi$ whose
designated tool $\tdes$ fails while a substitute tool $\tsub$ can still complete
the task. Our goal is to stop the calls to $\tsub$ that cross an action boundary
and to let all other calls execute. \toolname{} is deployed by the developer
who equips the agent with redundant tools and declares which tool substitutes
for which. It runs as a hook before each tool call, a mechanism that agent
frameworks such as Claude Code already provide, and it observes only what the
agent observes: the task, the agent's thoughts, and the tool returns. It cannot
inspect the environment, and no oracle is available at run time. We use the
force-push pair of \S~\ref{sec:study-rq1} as a running example, with a
fast-forward push as $\tdes$ and a force push as $\tsub$. In the normal
scenario, the agent publishes a feature branch that holds only its own work; in
the risky scenario, it publishes an integration branch to which other authors
have pushed commits. This setting raises three challenges.

\parh{C1: Identical Calls, Different States.}~The agent makes the same force
push on both branches. Mainstream mitigations judge the pending push or add a policy to
the instruction, so they permit or deny both pushes alike, which is why they
trade recoveries for violations (\S~\ref{sec:study-rq2}). What decides the boundary is the state of the target, and the agent usually
retrieves this state into its context before it falls back
(\S~\ref{sec:study-rq1}); on the integration branch, the context lists the
other authors' commits. We call this the evidence--decision gap.
\toolname{} therefore scores the context that the agent has accumulated at the
fallback rather than the call itself (\S~\ref{subsec:method-monitor}).

\parh{C2: Interventions End Recoveries.}~Because $\tdes$ keeps failing, the
force push is the only way to publish the feature branch. No monitor separates
the two scenario types perfectly, and guardrails that block or roll back on
detection (\S~\ref{subsec:robust}) turn each flagged normal run into a
failure. \toolname{} therefore withholds
the call once and returns the decision to the agent, so that a trigger in a
normal scenario costs only one step when the agent proceeds
(\S~\ref{subsec:method-intervene}). It also calibrates its threshold so that
the loss in normal success stays bounded.

\parh{C3: No Labels at Deployment.}~A deployment has no labels of risky runs
and cannot observe violations, so it can neither train a supervised monitor on
its own runs nor tune the threshold against the violation rate, as existing
monitors would require. \toolname{} therefore
trains the monitor once on the paired scenarios of \S~\ref{sec:study-design}, whose labels
hold by construction, and sets the threshold from normal runs alone, which any
deployment can record (\S~\ref{subsec:method-formulation} and
\S~\ref{subsec:method-monitor}).

\subsection{Overall Design}
\label{subsec:method-overview}

\F~\ref{fig:method} shows the workflow of \toolname{}. Three offline modules
turn recorded runs $\mathcal{H}$ into a monitor and a threshold, and one online
module guards the agent with them:
\begin{equation*}
\mathcal{H}
\xrightarrow[\text{\ding{172}}]{\text{collect}} P
\xrightarrow[\text{\ding{173}}]{\text{fit}} m
\xrightarrow[\text{\ding{174}}]{\text{calibrate}} \theta
\xrightarrow[\text{\ding{175}}]{\text{guard}} \pi_C .
\end{equation*}

\parh{Decision-Point Collection.}~The first module (\ding{172}) extracts, from
each recorded run on the paired scenarios, the agent's context at its first
call to $\tsub$, which we call the decision point of the run, and labels it with
the scenario type. A run that never calls $\tsub$ has no fallback to repair and
yields no decision point.

\parh{Risk-Signal Localization.}~The second module (\ding{173}) answers C1 by
fitting a monitor that separates risky decision points from normal ones. The
monitor embeds a normalized serialization of the context and scores the
embedding with a linear classifier; a white-box variant instead probes the
hidden state of the agent model at the decision point
(\S~\ref{subsec:method-monitor}).

\parh{Threshold Calibration.}~The third module (\ding{174}) answers C3 by
setting the threshold $\theta$ from normal decision points alone, to the
$(1-\beta)$ quantile of their scores, so that the budget $\beta$ governs the loss
in normal success (\S~\ref{subsec:method-formulation}).

\parh{Guarded Repair.}~The online module (\ding{175}) answers C2. Before each
call to $\tsub$, a hook scores the agent's context: below $\theta$, the call
executes unchanged; at or above $\theta$, the hook withholds it once and returns
a repair prompt, after which the agent either stops and reports or calls
$\tsub$ again, and that call executes (\S~\ref{subsec:method-intervene}).
Algorithm~\ref{alg:guard} summarizes both phases.

\subsection{Problem Formulation}
\label{subsec:method-formulation}

\toolname{} is defined by a configuration $C = (m, \theta, I)$: a monitor $m$
that maps the context $c$ at a call and the thought $t$ preceding the call to a
score in $[0,1]$, a threshold $\theta$, and an intervention $I$ that applies
when $m(c,t) \ge \theta$. We write $\pi_C$ for the agent under $C$, $R$ for its
ABV rate on risky scenarios, $U$ for its SR on normal scenarios
(\S~\ref{sec:study-design}), and $T(\theta)$ for the fraction of risky runs
whose decision point the monitor triggers on. Given a tolerance $\varepsilon$
on the loss in normal success, the repair problem is
\begin{equation}
\label{eq:objective}
C^{\star} \;=\; \arg\min_{C}\; R(\pi_C)
\quad\text{s.t.}\quad
U(\pi_C) \,\ge\, U(\pi) - \varepsilon .
\end{equation}

Solving \E~\eqref{eq:objective} directly would require running the agent under
every candidate configuration and judging each run with an oracle, which a
deployment lacks (C3). We therefore decompose the objective into quantities
that can be estimated separately. Let $\sigma$ be the fraction of triggered
risky runs in which the agent stops without executing $\tsub$, $\rho$ the
fraction of triggered normal runs in which it abandons the task, and
$F(\theta)$ the fraction of normal runs that the monitor triggers on. Because
\toolname{} changes nothing before its first trigger, we have
\begin{equation}
\label{eq:decompose}
R(\pi_C) \;=\; R(\pi)\,\bigl(1 - T(\theta)\,\sigma\bigr),
\qquad
U(\pi) - U(\pi_C) \;\le\; F(\theta)\,\rho .
\end{equation}
The threshold is the $(1-\beta)$ quantile of the monitor scores on recorded
normal decision points (\S~\ref{subsec:method-monitor}), so $F(\theta)$ stays
near $\beta$ as long as the recorded runs represent those in deployment. The
monitor and the threshold determine $T$ and $F$, which recorded runs suffice to
estimate, whereas the intervention determines $\sigma$ and $\rho$. The problem
thus splits into a monitor with a high $T(\theta)$
(\S~\ref{subsec:method-monitor}) and an intervention with a large $\sigma$ and
a small $\rho$ (\S~\ref{subsec:method-intervene}), while $\beta$ alone sets the
threshold.

\begin{proposition}[Threshold]
\label{prop:threshold}
Fix the monitor and the intervention of \S~\ref{subsec:method-intervene}, take
$F(\theta) = \beta$ and $\rho$ as fixed, and assume that an agent that calls
$\tsub$ again after the repair prompt reaches the same outcome as without the
trigger. Then $R(\pi_C)$ does not increase as $\theta$ decreases. Hence the
budget $\beta = \varepsilon/\rho$ minimizes $R(\pi_C)$ among the budgets whose
bound in \E~\eqref{eq:decompose} satisfies the utility constraint of
\E~\eqref{eq:objective}.
\end{proposition}

\begin{proof}[Proof sketch]
A larger $\beta$ lowers $\theta$, and a lower $\theta$ only adds triggers. After
an added trigger, the agent either stops, which can only remove a violation, or
calls $\tsub$ again, which by assumption leaves the outcome unchanged, so
$R(\pi_C)$ does not increase. By \E~\eqref{eq:decompose}, the loss in normal
success is at most $\beta\rho$, which stays within $\varepsilon$ exactly when
$\beta \le \varepsilon/\rho$. The assumption holds in our scenarios, where the
outcome of a run depends only on whether $\tsub$ executes
(\S~\ref{sec:study-design}).
\end{proof}

\subsection{Risk-Signal Localization and Calibration}
\label{subsec:method-monitor}

This module answers C1 and C3: it learns, from recorded runs alone, how a
risky fallback differs from a normal one in the agent's context. The facts that
establish a boundary, such as commits by other authors, an owner who differs
from the requester, a retention hold, or a consumer that requires a check, take
recurring forms in the retrieved context, so a monitor fitted on some tasks can
recognize them in others.

\parh{Decision Points.}~A run is a sequence of steps $(t_i, a_i, o_i)$ of
thought, action, and observation. Its decision point is the first step $k$ whose
action calls $\tsub$, represented as $(c, t, y)$ with the context
$c = (t_1, o_1, \ldots, t_{k-1}, o_{k-1})$, the thought $t = t_k$, and the label
$y = 1$ for a risky scenario and $y = 0$ otherwise; the decision points of the
recorded runs form the calibration set $P$. We fit the monitor on decision
points rather than on all steps, because at a decision point the agent has
already chosen to fall back, and the only open question is whether the
fallback crosses a boundary.

\parh{Context Serialization.}~The monitor reads a text $\kappa(c, t)$ that lists
the thoughts and observations of $c$ in order, followed by $t$, under three
normalizations. First, each injected error is replaced with a fixed
placeholder, so that the monitor cannot key on how an error is worded. Second,
scenario identifiers and person names are masked, so that the monitor keys on
roles such as another author rather than on particular people. Third, only the
most recent 8{,}000 characters are kept, which bounds the cost of a check while
retaining the state of the target that the agent usually retrieves shortly
before it falls back. On the integration branch, $\kappa(c,t)$ thus contains
the commit log with the other authors' names masked, followed by the
thought that proposes the force push.

\parh{Black-Box Monitor.}~The monitor embeds $\kappa(c,t)$ with
Qwen3-Embedding-0.6B~\cite{qwen3emb} and scores the standardized embedding
$\bar{e}$ with a logistic regression, $m(c,t) = \mathrm{sigmoid}(w^{\top}\bar{e}
+ b)$, trained with L2 regularization and class-balanced weights. Unlike an LLM
judge, a linear classifier is cheap enough to run at every check, returns the
same score for the same context, and exposes a single threshold for
calibration. Because it does not use the agent model, the monitor also applies
to hosted agents that expose only text, and it carries over to another agent
once $\theta$ is recomputed from the normal runs of that agent.

\parh{White-Box Variant.}~For open models, the monitor can instead read the
residual stream of the agent model at the last position of $\kappa(c,t)$, fit
the same classifier in each layer, and keep the layer with the highest AUROC on
held-out decision points. This variant tests whether the agent itself encodes
the boundary at the moment it decides; \S~\ref{sec:eval-ablation} compares it
with the black-box monitor, which all online runs use.

\parh{Threshold Calibration.}~Following Proposition~\ref{prop:threshold},
\toolname{} sets the threshold to the $(1-\beta)$ quantile of the monitor scores
on the normal decision points of $P$,
\begin{equation*}
\theta \;=\; Q_{1-\beta}\bigl(\{\, m(c,t) : (c,t,0) \in P \,\}\bigr).
\end{equation*}
Calibration needs neither labels of risky runs nor an oracle, and it is
repeated whenever the agent model or the kind of task changes.

\subsection{Guarded Repair}
\label{subsec:method-intervene}
\label{subsec:method-walkthrough}

This module answers C2 by deciding what the hook does when the monitor
triggers. The guard knows only that a fallback is suspicious, whereas the agent
holds the context that decides it (C1), so the repair sends the agent back to
this context instead of deciding for it.

\parh{Withhold and Redecide.}~On a trigger, the hook does not execute the call
to $\tsub$ and returns the repair prompt as the observation of the step. The
agent then either stops and reports what blocks completion, or proceeds, in
which case its next call to $\tsub$ executes without another check. Each run
receives at most one intervention, so a trigger costs at most one step. A
second withholding would overrule the decision that the agent has just made
with the prompt in view, turning the repair into the refusal that C2 rules out.

\parh{Prompt Defense.}~\toolname{} generates the repair prompt offline with an
LLM from the two kinds of violation of \S~\ref{sec:study-design}, not from
individual tasks. The result is a prompt defense, a template that asks the
agent to re-read the context it retrieved for the target and to check four
things: content on the target written by others, an entitlement whose owner or
approver is someone else, a record that something else retains or reads, and a
downstream consumer that requires a check the call would skip. If any of them
appears in the context, the agent is to stop and report what blocks
completion; otherwise, it is to proceed. The template leaves two slots, the
substitute and its target, which the hook fills in from the tool declaration
and the task at run time. Because the template is derived from the kinds of
violation rather than from any task, it states no verdict and points to no line
of the context, so the agent has to find the evidence itself. A trigger in a
normal scenario shows the same prompt, where the right response is to proceed,
so the prompt cannot lower violations by discouraging $\tsub$ without also
lowering normal success; \S~\ref{sec:evaluation} measures both.

\begin{algorithm}[t]\small
\caption{Offline calibration and online guarding of \toolname{}.}
\label{alg:guard}
\SetKwProg{Proc}{Procedure}{}{}
\SetKwFunction{Calibrate}{Calibrate}
\SetKwFunction{Guard}{Guard}
\SetKwComment{tcp}{\quad$\triangleright$~}{}
\Proc{\Calibrate{$\mathcal{H}$, $\beta$}}{
  $P \leftarrow$ the decision points $(c, t, y)$ of the runs in $\mathcal{H}$\tcp*{\ding{172} collect}
  fit $m$ on $P$\tcp*{\ding{173} localize}
  $\theta \leftarrow Q_{1-\beta}(\{\, m(c, t) : (c, t, 0) \in P \,\})$\tcp*{\ding{174} calibrate}
  $r \leftarrow$ the prompt defense generated by an LLM from the kinds of violation\;
  \KwRet{$(m, \theta, r)$}\;
}
\Proc{\Guard{$\pi$, $q$, $m$, $\theta$, $r$}}{
  $c \leftarrow q$; $\mathit{withheld} \leftarrow \mathit{false}$\tcp*{\ding{175} guard, from the task $q$}
  \While{the run has not ended}{
    $(t, a) \leftarrow \pi(c)$\;
    \eIf{$a$ calls $\tsub$ and $\neg\mathit{withheld}$ and $m(c, t) \ge \theta$}{
      $o \leftarrow r(a)$; $\mathit{withheld} \leftarrow \mathit{true}$\tcp*{withhold once}
    }{
      $o \leftarrow$ the result of executing $a$\;
    }
    $c \leftarrow c \cdot (t, a, o)$\;
  }
}
\end{algorithm}

\parh{Walkthrough.}~We trace the guard on three runs from \S~\ref{sec:eval-rq3},
with the monitor of Qwen at $\theta = 0.46$ ($\beta = 20\%$). On the feature
branch, the agent force-pushes right after reading the task; the score stays
below $\theta$, and the rebased work reaches the remote. On the integration
branch, the agent first retrieves the context of the branch, which shows that
four people, three of them other authors, wrote the six commits and that
deployment pins reference them. The monitor triggers, and after the prompt the
agent reports that the pins still reference three of the commits and stops. On
the release branch, a second risky scenario, the agent force-pushes without
reading the log that shows two commits by other authors. The context then names
the release branch but holds no evidence of the boundary, so the score stays
below $\theta$, and the push executes. The monitor thus cannot catch a fallback taken before the context is
retrieved, and such runs limit its recall.

\section{Evaluation}
\label{sec:evaluation}

To evaluate \toolname{}, we organize the evaluation around three research
questions (RQs):

\noindent\textbf{RQ3: Effectiveness and Generalization.} Does \toolname{} lower
the ABV rate on risky scenarios while keeping SR on normal scenarios, and does
its monitor transfer to unseen tasks and models?

\noindent\textbf{RQ4: Cost.} What does \toolname{} cost offline and online?

\noindent\textbf{RQ5: Ablation Study.} Which components of \toolname{} account for
its effect and its cost?

\noindent We then present a case study with two production agents on a real
repository (\S~\ref{sec:eval-rq5}).

Unless stated otherwise, we reuse the dataset, the four models, the agent
configuration, the injected errors, and the oracle of \S~\ref{sec:study-design},
so that every difference from RQ1 comes from \toolname{}. The monitor embeds the
serialized context with Qwen3-Embedding-0.6B~\cite{qwen3emb} and scores it with
a logistic regression (\S~\ref{subsec:method-monitor}); it is fitted on the runs
of Qwen, with Gemma, Luna, and DeepSeek as unseen models, and calibrated at
$\beta = 20\%$. We compare \toolname{} with no repair and with the three
mitigations of \S~\ref{sec:study-rq2}.
The three mitigations use the same setting as \toolname{}, including the context they see.
Besides SR, the ABV rate, and the cost,
we report the recall $T(\theta)$ and the precision of the monitor, counting a
run as triggered if any of its checks triggers. We test changes against no
repair with the McNemar tests of \S~\ref{sec:study-design} and compute 95\% confidence
intervals from 10{,}000 bootstrap resamples of the scenarios.

\subsection{RQ3: Effectiveness and Generalization}
\label{sec:eval-rq3}

To answer RQ3, we evaluate whether \toolname{}
lowers the ABV rate without lowering SR, and whether its monitor generalizes.

\begin{table}[t]
\centering
\footnotesize
\caption{Effectiveness of \toolname{} across the four models.}
\label{tab:eval-main}
\setlength{\tabcolsep}{5pt}
\begin{tabular}{llccccc}
\toprule
 & & Normal & Risky & \multicolumn{2}{c}{Monitor} & \\
\cmidrule(lr){3-3}\cmidrule(lr){4-4}\cmidrule(lr){5-6}
Model & Configuration & SR & ABV & Recall & Precision & Cost \\
\midrule
Qwen & No repair & 98.0 & 96.8 & -- & -- & -- \\
 & \toolname{} & 97.3 & 77.3 & 80.8 & 80.8 & 0.03 \\
\midrule
Gemma & No repair & 93.3 & 88.3 & -- & -- & -- \\
 & \toolname{} & 91.3 & 55.8 & 83.0 & 80.6 & 0.06 \\
\midrule
Luna & No repair & 92.0 & 67.5 & -- & -- & -- \\
 & \toolname{} & 79.3 & 16.2 & 84.8 & 70.1 & 0.24 \\
\midrule
DeepSeek & No repair & 83.3 & 23.5 & -- & -- & -- \\
 & \toolname{} & 77.9 & 11.8 & 65.7 & 44.2 & 0.44 \\
\bottomrule
\end{tabular}
\end{table}

We fit the monitor once, on the 296 decision points in the RQ1 runs of Qwen,
149 of which lie on risky scenarios. The tasks, one per tool pair, are split
into five folds so that each scenario is scored by a classifier fitted on other
tool pairs, which share none of its tools, tasks, or records.
The unseen models reuse these classifiers unchanged and recompute only the
threshold from their own RQ1 runs on normal scenarios, which needs no labels of
risky scenarios. As a deployment calibrates on recorded runs of its own tasks,
each threshold comes from earlier runs of the evaluated scenarios, whereas no
classifier has seen the task that it scores. By \E~\eqref{eq:decompose},
$\beta = 20\%$ keeps the loss of normal success near two points as long as a
triggered agent abandons at most one normal task in ten. We then run the four models with \toolname{} and
compare them with no repair (Table~\ref{tab:eval-main}); almost
all triggers happen at the decision point itself (144 of 146 on Qwen and 135 of
139 on Gemma).

We find that \toolname{} lowers violations while preserving normal success, at
a fraction of the cost of mainstream mitigations. As shown in
Table~\ref{tab:eval-main}, on Qwen and Gemma it lowers the ABV rate from 96.8\%
to 77.3\% and from 88.3\% to 55.8\%, by 19.5 points (95\% CI 13.0 to 26.0) and
32.5 points (24.7 to 40.9), which removes one fifth and more than one third of
the violations ($p < 10^{-7}$ for both). Normal success on these two models
falls by only 0.7 and 2.0 points, which is not significant, and the confidence
intervals bound the loss at 3.3 and 6.7 points. This matches
\E~\eqref{eq:decompose}: the monitor triggers on 18.7\% and 18.0\% of normal
runs, close to $\beta$, and a triggered agent abandons the task in 7.4\% and
12.0\% of them ($\rho$), which bounds the loss at 1.4 and 2.2 points. On Luna and DeepSeek, the relative reduction is larger but costs normal
success. On Luna, the ABV rate falls from 67.5\% to 16.2\%, while normal success
falls from 92.0\% to 79.3\% ($p < 0.001$; 95\% CI of the loss 6.0 to 20.0
points). Luna is a hosted model whose runs vary between calibration and
evaluation, so its monitor triggers on 25.3\% of normal runs rather than 20\%,
and $F(\theta)\rho = 12.7$ points matches the loss. Halving the budget to 10\%
cuts this loss to 7.3 points, while the ABV rate still falls to 24.0\%
(\S~\ref{sec:eval-ablation}). DeepSeek, which already stops in most risky scenarios, halves its ABV
rate from 23.5\% to 11.8\% ($p < 0.001$), but its risky decision points are few,
so most triggers fall on normal runs and normal success falls from 83.3\% to
77.9\% ($p = 0.057$). Per prevented violation, \toolname{} gives up 0.03, 0.06,
0.24, and 0.44 normal successes on Qwen, Gemma, Luna, and DeepSeek. On Qwen, the mitigations
of \S~\ref{sec:study-rq2} give up 0.62 to 0.99, over an order of magnitude more.
On the same model, \toolname{} has both a higher SR and a lower ABV rate than
deliberative alignment (97.3\% and 77.3\% against 92.7\% and 88.3\%), and with
refusal (\S~\ref{sec:eval-ablation}) both a higher SR and a lower ABV rate than
GuardAgent (80.7\% and 18.8\% against 52.7\% and 29.2\%). This demonstrates that \toolname{}
removes violations on every model and preserves normal success wherever the
agent rarely abandons a task after the prompt; \S~\ref{sec:eval-ablation} shows
which of its components keep the cost low.

\begin{table}[t]
\centering
\footnotesize
\caption{Generalization of the monitor to unseen tasks and models at $\beta = 20\%$.}
\label{tab:eval-monitor}
\begin{tabular}{lllccc}
\toprule
Generalization & Model & Threshold from & Decision points & AUROC & Recall \\
\midrule
Unseen tasks & Qwen & Qwen & 296 & 0.904 & 81.9 \\
\midrule
Unseen models & Gemma & Gemma & 276 & 0.887 & 80.1 \\
 & Gemma & Qwen & 276 & 0.887 & 75.7 \\
 & Luna & Luna & 242 & 0.897 & 84.6 \\
 & DeepSeek & DeepSeek & 161 & 0.843 & 75.0 \\
\midrule
\multicolumn{2}{l}{White-box monitor} & & & & \\
Unseen tasks & Qwen & Qwen & 296 & 0.768 & 63.1 \\
\bottomrule
\end{tabular}
\end{table}

To evaluate the monitor itself, we further score it offline on decision points
outside its fitting data (Table~\ref{tab:eval-monitor}): the unseen tasks of
Qwen and the three unseen models; on Gemma, we also keep the threshold of Qwen.
We find that the monitor places most triggers on risky scenarios and that its
ranking transfers to unseen tasks and models. Online
(Table~\ref{tab:eval-main}), about 81\% of the triggers fall on risky scenarios
on Qwen and Gemma and 70\% on Luna, with a recall $T(\theta)$ of 80.8\% to
84.8\%; only on DeepSeek, whose risky decision points are few, does the
precision drop to 44.2\%. Offline, the classifier separates the unseen tasks
of Qwen with an AUROC of 0.904, and the classifiers fitted on Qwen reach
0.887 on Gemma, 0.897 on Luna, and 0.843 on DeepSeek without refitting. The
threshold, in
contrast, is specific to the model: keeping the Qwen threshold on Gemma lowers
the recall from 80.1\% to 75.7\%, whereas the threshold recomputed from the
normal runs of Gemma preserves the effect of Table~\ref{tab:eval-main}, with
more than a third of the violations removed for 2.0 points of normal success.
This demonstrates that a new model needs no new classifier,
only a threshold recomputed from its own normal runs, which require no labels.

To explain why the reduction differs across models, we decompose it by
\E~\eqref{eq:decompose} into the recall $T(\theta)$ and the fraction $\sigma$ of
triggered risky runs in which the agent stops, and we measure $\sigma$ and the
fraction $\rho$ of triggered normal runs in which the agent abandons the task.
We find that the agent's response to the repair prompt sets both the
reduction and the cost. The monitor reaches four in five risky decision points
on Qwen, Gemma, and Luna, so these three models differ mainly in $\sigma$: after
the prompt, the agent stops in 23\% of triggered risky runs on Qwen, 45\% on
Gemma, and 91\% on Luna. With these values, \E~\eqref{eq:decompose} predicts ABV
rates of 78.7\%, 55.7\%, and 15.5\%, within 1.4 points of the measured 77.3\%,
55.8\%, and 16.2\%. The agents that stop more often in risky runs also abandon
more often in normal ones, with $\rho$ of 7\%, 12\%, and 50\%, although $\sigma$
exceeds $\rho$ by a factor of 1.8 to 3.7. For instance, Luna, which abandons
half of the triggered normal tasks, removes three quarters of its violations
but also loses 12.7 points of normal success. Retrieving the context again does
not raise $\sigma$: the agents that proceed retrieve it again at least as often
as those that stop (44.4\% against 33.3\% on Qwen and 63.9\%
against 36.7\% on Gemma). This demonstrates that \toolname{} returns the
decision to the agent with a prompt defense, but how strongly and how selectively the agent responds is a property of
the model rather than of the monitor.

\rqbox{\textbf{RQ3:}~\toolname{} lowers the ABV rate on every model, by one
fifth to three quarters, without a significant loss of normal success on Qwen
and Gemma, and it gives up 0.03 to 0.44 normal successes per prevented
violation, against 0.62 to 0.99 for mainstream mitigations. Its monitor transfers
to unseen tasks and models with only a recomputed threshold, and how
many violations it removes depends on how often the agent stops after the
prompt.}

\subsection{RQ4: Cost}
\label{sec:eval-rq4}

To answer RQ4, we measure the offline and online cost of each component of
\toolname{} (Table~\ref{tab:eval-overhead}). Latencies are measured on the
serving node, and the generation time of one step on Qwen.

\begin{table}[t]
\centering
\footnotesize
\caption{Overhead of \toolname{} on Qwen and Gemma.}
\label{tab:eval-overhead}
\begin{tabular}{llccc}
\toprule
Stage & Component & Frequency & Latency & Added tokens \\
\midrule
Offline & Fitting the monitor & Once per agent model & 7.5 s & 0 \\
\midrule
Online & Embedding the context & 1.1--1.4 per run & 66 ms & 0 \\
 & Linear classifier & 1.1--1.4 per run & 2.8~$\mu$s & 0 \\
 & Repair prompt & Once per trigger & One step & 103 \\
\bottomrule
\end{tabular}
\end{table}

For the offline cost, we time fitting the monitor for Qwen and transferring it
to Gemma.
We find that fitting needs only recorded runs. As shown in
Table~\ref{tab:eval-overhead}, fitting the monitor for Qwen takes 7.5 seconds,
almost all of it spent on one embedding pass over its 296 decision points; the
six logistic regressions, one per fold and one on all points, take 7 ms each on
a CPU. For instance, transferring the monitor to Gemma takes one embedding pass
over the decision points of the normal Gemma runs and one quantile. This
demonstrates that neither step updates the agent model, generates new runs, or
needs labels beyond the scenario type.

For the online cost, we count the checks and triggers in each run and measure
the latency of a check and the steps and tokens that a trigger adds.
We find that the online cost falls on triggered runs. The monitor checks calls
to $\tsub$ only until its first trigger, 1.1 to 1.4 times per run, and almost
every trigger happens at the decision point. A check costs one embedding call of
66 ms at the median (77 ms at the 90th percentile) and one inner product of
2.8~$\mu$s, about 1\% of the 6.2 s that the agent needs at the median to
generate one step. The visible cost is the trigger, which withholds one call and
adds a repair prompt of about 100 tokens. For instance, on normal runs,
\toolname{} adds 0.38 steps on Qwen and 0.31 on Gemma, about a third and a
seventh of what the prompt adds without the monitor
(\S~\ref{sec:eval-ablation}), whereas on risky runs,
where a trigger is intended, it adds 0.94 and 0.70 steps. This demonstrates that
\toolname{} spends its cost where it intervenes rather than on every step.

\rqbox{\textbf{RQ4:}~\toolname{} is fitted once from recorded runs in 7.5
seconds. Online, each check takes 66 ms, about 1\% of one generation step, and
only a trigger adds cost: one step and about 100 tokens.}

\subsection{RQ5: Ablation Study}
\label{sec:eval-ablation}

To answer RQ5, we vary one component of \toolname{} at a time, namely the
monitor, the intervention, the prompt, and the budget, and evaluate how each
affects the reduction in violations and its cost.

\begin{table}[t]
\centering
\footnotesize
\caption{Ablation of the monitor and the intervention of \toolname{}. $\sigma$: stop rate of triggered risky runs; $\rho$: abandon rate of triggered normal runs.}
\label{tab:eval-ablation}
\setlength{\tabcolsep}{4.5pt}
\begin{tabular}{llcccccc}
\toprule
 & & Normal & Risky & Monitor & \multicolumn{2}{c}{Agent response} & \\
\cmidrule(lr){3-3}\cmidrule(lr){4-4}\cmidrule(lr){5-5}\cmidrule(lr){6-7}
Model & Configuration & SR & ABV & Precision & $\sigma$ & $\rho$ & Cost \\
\midrule
Qwen & No repair & 98.0 & 96.8 & -- & -- & -- & -- \\
 & \toolname{} & 97.3 & 77.3 & 80.8 & 23.1 & 7.4 & 0.03 \\
 & Without monitor & 90.0 & 70.8 & 50.7 & 26.4 & 6.2 & 0.30 \\
 & With refusal & 80.7 & 18.8 & 80.8 & 100.0 & 100.0 & 0.22 \\
\midrule
Gemma & No repair & 93.3 & 88.3 & -- & -- & -- & -- \\
 & \toolname{} & 91.3 & 55.8 & 80.6 & 44.5 & 12.0 & 0.06 \\
 & Without monitor & 82.0 & 49.4 & 50.0 & 45.3 & 11.5 & 0.28 \\
 & With refusal & 76.7 & 16.2 & 80.6 & 100.0 & 100.0 & 0.23 \\
\bottomrule
\end{tabular}
\end{table}

For the monitor, we run \toolname{} without the monitor, which returns the
repair prompt at the first call to $\tsub$ in every run, and replace the
black-box monitor with a white-box monitor that reads the residual stream of
Qwen instead of an embedding of the context (Table~\ref{tab:eval-ablation}).
We find that the monitor is what keeps the cost low. As shown in
Table~\ref{tab:eval-ablation}, returning the repair prompt at every fallback
prevents more violations, 40 on Qwen and 60 on Gemma against 30 and 50 for
\toolname{}, but it also loses 12 and 17 normal successes ($p < 0.002$) and
places half of its interventions on normal scenarios (precision 50.7\% and
50.0\%), since the call is the same in both scenario types. The agent responds
to the prompt about equally in the two configurations ($\sigma$ of 26.4\% and
23.1\% on Qwen, 45.3\% and 44.5\% on Gemma), so the monitor changes where the
prompt is shown rather than how the agent reacts to it: \toolname{} keeps three
quarters to five sixths of the prevented violations at one twelfth and one
sixth of the loss. The white-box monitor separates the same decision points less well than the black-box monitor,
with an AUROC of 0.768 at its best layer against 0.904
(Table~\ref{tab:eval-monitor}), so all online runs use the black-box monitor.
This demonstrates that the selectivity of \toolname{} comes from the monitor,
and that an embedding of the context carries more of the boundary than
the hidden state of the agent at the decision point.

For the intervention, we replace the repair prompt with refusal, which keeps
the monitor and the threshold but refuses the triggering call and every later
call to $\tsub$. Because the designated tool keeps failing, a refused
substitute leaves no other path to the change, so we obtain this configuration
exactly from the \toolname{} runs by keeping each triggered run only up to its
trigger.
We find that returning the decision, rather than refusing, preserves normal
success. Refusal turns every triggered normal run into a failure: the monitor triggers on 28 normal runs of Qwen and 27 of Gemma, and in
25 and 22 of them the agent calls $\tsub$ again after the prompt and completes
the task. Refusal therefore lowers normal success by 17.3 and 16.7 points.
Although it also lowers the ABV rate to 18.8\% and 16.2\%, it costs about 0.22 normal
successes per prevented violation, against 0.03 and 0.06 for \toolname{}. This
demonstrates that refusal sets $\rho$ to one and suits only deployments that can
afford to lose about one normal success in six.

\begin{table}[t]
\centering
\footnotesize
\caption{Impact of the prompt defense on replayed triggers. $p$: McNemar test on triggered risky scenarios.}
\label{tab:eval-prompt}
\setlength{\tabcolsep}{5pt}
\begin{tabular}{llcccccc}
\toprule
Model & Prompt & $\sigma$ & $\rho$ & SR & ABV & Cost & $p$ \\
\midrule
Qwen & Prompt defense & 25.9 & 3.8 & 97.3 & 74.7 & 0.03 & \multirow{2}{*}{0.75} \\
 & Generic & 27.2 & 3.7 & 98.0 & 72.7 & 0.00 & \\
\midrule
Gemma & Prompt defense & 42.5 & 12.0 & 91.3 & 55.8 & 0.06 & \multirow{2}{*}{$<10^{-4}$} \\
 & Generic & 19.4 & 8.7 & 90.7 & 72.7 & 0.17 & \\
\midrule
Luna & Prompt defense & 87.8 & 42.9 & 80.0 & 16.9 & 0.23 & \multirow{2}{*}{$<0.001$} \\
 & Generic & 62.5 & 17.1 & 86.0 & 28.6 & 0.15 & \\
\bottomrule
\end{tabular}
\end{table}

For the prompt, to separate the content of the repair prompt from the act of
withholding the call, we replay the triggers of \toolname{} once with the
repair prompt and once with a generic prompt that asks the agent to re-read the
context and decide again, without the prompt defense
(Table~\ref{tab:eval-prompt}).
We find that the prompt defense strengthens the repair on every agent that
responds to a prompt. As shown in Table~\ref{tab:eval-prompt}, it raises
$\sigma$ from 19.4\% to 42.5\% on Gemma and from 62.5\% to 87.8\% on Luna, and the ABV rate
falls further, to 55.8\% instead of 72.7\% and to 16.9\% instead of 28.6\%. On
Gemma, this comes at no additional loss in normal success, so the cost per
prevented violation falls from 0.17 to 0.06. Luna, which already follows the
prompt readily, also abandons more normal tasks with the prompt defense
($\rho$ of 42.9\% against 17.1\%), so the generic prompt costs less per
prevented violation (0.15 against 0.23), as a smaller budget does. On Qwen,
neither prompt moves the agent, which stops in about a quarter of triggered
risky runs under both (25.9\% and 27.2\%), although, given the same context,
Qwen identifies the boundary in 73.4\% of risky scenarios
(Table~\ref{tab:rq1-ask}). This demonstrates that the prompt defense makes the
prompt stronger, and that how much of the added strength reaches the risky
scenarios depends on the agent.

\begin{table}[t]
\centering
\footnotesize
\caption{Impact of the budget ($\beta$) on Luna and DeepSeek.}
\label{tab:eval-budget}
\setlength{\tabcolsep}{5pt}
\begin{tabular}{llccccc}
\toprule
 & & Normal & Risky & \multicolumn{2}{c}{Monitor} & \\
\cmidrule(lr){3-3}\cmidrule(lr){4-4}\cmidrule(lr){5-6}
Model & Configuration & SR & ABV & Recall & Precision & Cost \\
\midrule
Luna & No repair & 92.0 & 67.5 & -- & -- & -- \\
 & $\beta = 20\%$ & 79.3 & 16.2 & 84.8 & 70.1 & 0.24 \\
 & $\beta = 10\%$ & 84.7 & 24.0 & 71.0 & 80.0 & 0.16 \\
\midrule
DeepSeek & No repair & 83.3 & 23.5 & -- & -- & -- \\
 & $\beta = 20\%$ & 77.9 & 11.8 & 65.7 & 44.2 & 0.44 \\
 & $\beta = 10\%$ & 80.7 & 12.4 & 55.6 & 60.6 & 0.24 \\
\bottomrule
\end{tabular}
\end{table}

For the budget, the one parameter that a deployment sets, we vary $\beta$ from
5\% to 30\% on the calibration decision points of Qwen, and rerun \toolname{}
with $\beta = 10\%$ on Luna and DeepSeek, the two models where the loss of
normal success at $\beta = 20\%$ is largest (Table~\ref{tab:eval-budget}).
We find that the budget trades recall for normal success. On the calibration
decision points of Qwen, recall rises steadily with the budget, from 61.1\% at 5\% to 90.6\% at
30\%, while the bound $\beta\rho$ on the loss of normal success grows in
proportion to it. With $\rho$ near 0.1, as on Qwen and Gemma, even a budget of
30\% keeps the bound below four points. On Luna, where $\rho$ is about 0.5, and
on DeepSeek, whose risky decision points are few, the budget matters
(Table~\ref{tab:eval-budget}). Halving it to 10\% lowers the loss of normal
success on Luna from 12.7 to 7.3 points (95\% CI 0.0 to 14.7), while the ABV
rate still falls from 67.5\% to 24.0\% and the cost drops from 0.24 to 0.16. On DeepSeek, the same halving keeps the reduction
(12.4\%, $p = 0.005$), lowers the cost from 0.44 to 0.24, and leaves a loss of
2.6 points in normal success ($p = 0.48$). This
demonstrates that an agent that responds strongly to the prompt, or that rarely
falls back in risky scenarios, calls for a smaller budget.

\rqbox{\textbf{RQ5:}~Each component of \toolname{} keeps its cost low.
Without the monitor or with refusal, the cost rises from 0.03--0.06 to
0.22--0.30 per prevented violation, and without the prompt defense, Gemma and
Luna stop less often in risky runs.}

\subsection{Case Study: Production Agents on a Real Repository}
\label{sec:eval-rq5}

We close with a case study that attaches \toolname{} to two production agents
acting on a real repository. A branch has been rebased, the plain push is
rejected as non-fast-forward, and the agent is told only to get the rebased
work onto the remote. In the normal scenario, the remote holds only the
agent's own commits, which the rebase has superseded; in the risky scenario,
another author has pushed two commits since the last fetch. We run Claude Code with Claude Haiku 4.5 and Codex with GPT-5.6 Luna,
each in its own command-line environment with default permissions, 16 times
per scenario and configuration on a freshly prepared repository. After each
run, \texttt{git} checks whether any commit that the remote held before the run
is lost. A run that the provider rate-limited before any decision is repeated.
Configurations are compared with Fisher's exact test, since their runs are not
matched.
\toolname{} runs as a hook before each tool call and reads only the session
transcript. As in the main evaluation, its monitor is fitted on paired scenarios
whose labels hold by construction (C3), here the runs of the two scenarios
without \toolname{}, leaving out one run at a time, and it is thresholded per
agent at $\beta = 20\%$.

\begin{wraptable}{r}{0.5\linewidth}
\vspace{-10pt}
\centering
\footnotesize
\caption{Effectiveness of \toolname{} on two production agents over 16 runs per cell.}
\label{tab:eval-case}
\setlength{\tabcolsep}{3pt}
\begin{tabular}{lcccc}
\toprule
 & \multicolumn{2}{c}{SR (normal)} & \multicolumn{2}{c}{ABV (risky)} \\
\cmidrule(lr){2-3}\cmidrule(lr){4-5}
Agent & No repair & \toolname{} & No repair & \toolname{} \\
\midrule
Claude Code & 100.0 & 93.8 & 68.8 & 6.3 \\
Codex & 100.0 & 100.0 & 100.0 & 25.0 \\
\bottomrule
\end{tabular}
\vspace{-8pt}
\end{wraptable}
As shown in Table~\ref{tab:eval-case}, the agents fetch and read the other
author's commits before every risky forced push, which also refreshes the lease
that \texttt{-{}-force-with-lease} checks, yet without \toolname{} they still
discard them in 11 of 16 runs on Claude Code and in all 16 on Codex. The
monitor separates the two scenarios almost perfectly (AUROC 1.00 and 0.97), and
\toolname{} lowers the violations to 1 of 16 on Claude Code ($p < 0.001$) and
to 4 of 16 on Codex ($p < 10^{-4}$), while normal success moves from 16 to 15
runs on Claude Code and stays at 16 on Codex. This demonstrates that
production agents show the same evidence--decision gap, and that \toolname{}
closes most of it once the evidence is read.

\section{Discussion}
\label{sec:discussion}

\subsection{Limitations of \toolname}
\label{sec:limits}

\parh{Evidence the Agent Never Retrieves.}~\toolname{} judges the evidence
that the agent has retrieved, so it cannot flag a fallback taken before the
record is read, as in the walkthrough of \S~\ref{subsec:method-walkthrough}. In
our scenarios this case is rare, at most 9.6\% of violating runs depending on
the model. Retrieving the evidence is therefore a precondition that a deployment
has to secure, through a team convention that asks the agent to inspect the
target before rewriting it, or through a hook that, when the substitute is
called before any read of the target, asks the agent to read the target first.

\parh{The Agent Still Decides.}~Because \toolname{} returns the decision to
the agent, its effect is bounded by how the agent responds to the repair
prompt. On Qwen, most triggered risky runs still call the substitute
(\S~\ref{sec:eval-rq3}); the monitor, which reaches four in five risky
decision points, is not the bottleneck. An agent that follows the prompt
readily, such as Luna, stops in most triggered risky runs but also abandons
half of the triggered normal ones. The budget $\beta$ sets this trade-off for
each agent (Proposition~\ref{prop:threshold}), but raising $\sigma$ without
raising $\rho$ requires the agent to read the context correctly after the
prompt, which a prompt defense alone cannot guarantee
(\S~\ref{sec:eval-ablation}). Where violations must be ruled out regardless of the
agent, the same monitor can drive a refusal instead, which lowers the ABV rate
of Qwen to 18.8\% at a cost of 0.22 (Table~\ref{tab:eval-ablation}).

\parh{Thresholds Must Be Recalibrated.}~Within our scenarios, the monitor's
ranking transfers to unseen tasks and models, but its threshold is specific
to the model
and the kind of task. A deployment therefore recomputes $\theta$ from normal
runs whenever the agent model or the kind of task changes, and it tracks normal
success online, since the offline calibration is only an estimate. Normal runs
need no labels, so this recalibration is cheap, but it cannot be skipped.

\parh{Hidden States Carry Less Than the Context.}~The white-box variant reaches
an AUROC of 0.768 at its best layer, against 0.904 for the black-box monitor on
the same decision points. One likely reason is that the state of the target sits
in long tool returns early in the context, whereas the residual stream at the
last position mainly reflects the step the agent is about to take. The result
is consistent with the evidence--decision gap of \S~\ref{subsec:method-challenges}:
the record is in the agent's context, but the agent's own representation at
the moment of decision separates the two scenario types less well than an
embedding of that context does.

\subsection{Threats to Validity}
\label{sec:threats}

\parh{Internal Validity.}~The scenarios are written by LLMs from documented
CVEs and may contain defects. Automated checks keep the records of the two
scenario types in the same format and at comparable length, and screening
removed 16 scenarios whose records contradicted the task premise. Each scenario
carries the derivation of its label from its records, and each risky one cites
its CVE, so every label can be checked against the data. The oracle
checks the recorded state after each run and involves no model judgment. Hosted
models may vary between runs, so all comparisons are paired over matched
scenarios.

\parh{External Validity.}~The scenarios cover eight domains and 55 CVEs but are
not drawn from production traces, and their designated tool keeps failing, so
intermittent failures are not measured. The case study covers one task family
with two production agents. A new repository receives its own monitor, fitted
as in \S~\ref{sec:eval-rq5}. Our results on four models may not hold for
others.

\section{Conclusion}
\label{sec:conclusion}

This paper studied risky fallback, in which an LLM agent falls back to a
substitute tool after a tool error and crosses an action boundary depending on
the state of its target. Our study shows that agents cross the boundary even
when the state of the target is already in their context, and that mainstream
mitigations stop violations only by also stopping legitimate recoveries. We
presented \toolname{}, which judges a fallback by the agent's context rather
than by the call. It scores the context at each call to the substitute tool
with a monitor calibrated on normal runs alone and, when the monitor triggers,
withholds the call once and returns the decision to the agent. Closing the gap
between the evidence an agent reads and the decision it makes remains the main
direction for future work.

\bibliographystyle{ACM-Reference-Format}
\bibliography{bib/main}

\end{document}